\documentclass[twocolumn,showpacs,preprintnumbers,amsmath,amssymb,floatfix]{revtex4}

\usepackage{graphicx}
\usepackage{dcolumn}
\usepackage{bm}

\newcommand{\beq}{\begin{equation}}
\newcommand{\eeq}{\end{equation}}
\newcommand{\bey}{\begin{eqnarray}}
\newcommand{\eey}{\end{eqnarray}}

\begin{document}


\title{Anisotropic stellar models admitting conformal motion}

\author{Ayan Banerjee}
 \email{ayan_7575@yahoo.co.in}
\affiliation {Department of Mathematics, Jadavpur University,
 Kolkata-700032, India}
\author{Sumita Banerjee}
 \email{banerjee.sumita.jumath@gmail.com}
\affiliation {Department of Mathematics, Budge Budge Institute of Technology,
 West Bengal 700137, India}
\author{Sudan Hansraj}
 \email{hansrajs@ukzn.ac.za}
\affiliation {Astrophysics and Cosmology Research Unit, School of Mathematics,
Statistics and Computer Science, University of KwaZulu-Natal,
Private Bag X54001, Durban 4000, South Africa}
\author{Ali Ovgun}
 \email{ali.ovgun@emu.edu.tr}
\affiliation {Physics Department , Eastern Mediterranean University,
Famagusta, Northern Cyprus, Mersin 10, Turkey}

\date{\today}

\begin{abstract}
We address the problem of finding static and spherically
symmetric anisotropic compact stars in general relativity that admit conformal motions. The study is framed in the language of $f(R)$ gravity theory in order to expose opportunity for further study in the more general theory.
Exact solutions of compact stars are found under the assumption that
spherically symmetric spacetimes which admits conformal motion with matter distribution
 anisotropic in nature. In this work, two cases have been studied for the existence of such
solutions: first, we consider the model given by f(R) = R and then f(R) = aR+b. Finally, specific characteristics
and physical properties have been explored by analytically along with graphical representations
for conformally symmetric compact stars in f(R) gravity.

\end{abstract}


\maketitle

\section{Introduction:}
From the time of Sir Isaac Newton our understanding of the nature of gravity has
advanced but mysteries in physics still remain. The process is still continuing after
the first formulation of Einstein's theory of General Relativity (GR) \cite{Einstein}, which is
one of the most successful fundamental theories of gravity in modern physics. Despite its success, many extensions
of the original Einstein equations have been investigated to accommodate present observational data
on both astrophysical and cosmological scales. Observational data has confirmed
that the Universe is undergoing a phase of accelerated expansion. Direct support is provided by the
high precision data from Type Ia supernovae \cite{Supernova}, the Cosmic Microwave Background (CMB) anisotropies
\cite{Dunkley}, baryonic acoustic oscillations \cite{Eisenstein} and from gravitational lensing \cite{Contaldi}.
In particular, within the context of General Relativity,  the observed cosmic acceleration of the Universe cannot be explained without resorting to additional exotic matter components (such as Dark Energy (DE) or dark mater (DM)). Until now, no consistent model of dark energy has been proposed which can yield a convincing  self-consistent picture of the observed Universe, both in the field of fundamental physics and in that of astrophysics. Currently experimental tests for DE and DM are under design and it is believed that with precision instrumentation such as the Square Kilometre Array radio telescopes such ideas will either be confirmed or ruled out. For this reason modifications of General Relativity have been proposed and are currently receiving a great deal of attention.

 Alternative gravity theories generally appear to exhibit  good agreement between theory and observation.
and the introduction of additional geometrical degrees of freedom may assist us in abandoning the concepts
of DE or DM.   Many alternative theories have been proposed
such as f (R) gravity \cite{Capozziello}, scalar-tensor theories \cite{Setare}, braneworld models \cite{Dvali},
Gauss-Bonnet gravity \cite{Nojiri}, Galileon gravity \cite{Nicolis} have gained a lot of attention
in recent years. Additionally Ellis \cite{ellis1, ellis2} revived an idea proposed by Weinberg \cite{wein}, that of trace-free Einstein gravity, which has also gone by the name unimodular \cite{unimod1,unimod2} gravity. In this framework an attempt was made to rectify the discrepancy in the value of the cosmological constant as predicted by quantum gravity in contrast with what is actually observed. Dadhich \cite{dad-hans} has also presented strong arguments for using the Lovelock polynomial terms as the generator of the Lagrangian density. The remarkable feature of the Lovelock theory is that while it is polynomial in the Riemann tensor, Ricci tensor and Ricci scalar, it nevertheless results in at most second order equations of motion. Moreover, the zeroth case corresponds to the cosmological constant while the linear case yields the familiar Einstein gravity theory. The effects of the higher curvature terms are only felt in dimensions greater than 4 so that the theory is a natural generalisation of GR to higher dimensions. Investigations into higher dimensional gravity is strongly motivated by string theoretic ideas which are supposed to be compatible with gravity theories. It should be mentioned here that
many of these modifications have provided a number of significant results in
cosmology \cite{Maeda, Ananda} as well as astrophysics \cite{Capozziello}.

In the present article we have endeavoured to frame the core problem in the language and formalism of the f(R) theory although the specific cases under consideration reduce to the standard Einstein theory which contains an action principle that is linear in the Ricci scalar.  The reason for this choice
of formalism is that it opens the way for further investigations of more general functional forms of the
Ricci scalar such as the Starobinksy model  in subsequent work.
In f(R) gravity theories  the Einstein-Hilbert action is generalized
with a function of the Ricci scalar R, (see the reviews \cite{Faraoni,Buchdahl}) have also
been extensively analyzed recently. Earlier interest in f(R) theories occurred
in connection with the early Universe \cite{Faraoni}, and a number
of viable f(R) models have proposed that satisfy the observable constraints - see e.g., Refs. \cite{Amendola}.
It is noted that f(R) theory was contemplated long ago by Buchdahl \cite{Buchdahl}
who introduced f(R) gravity using non-linear Lagrangians.
One of the main reasons for increased interest in f(R) gravity was motivated by inflationary scenarios, in
the Starobinsky model, where f(R) = R-$\Lambda$+$\alpha R^2$ gravity are considered in \cite{Starobinsky}.
The Cosmology of generalized modified gravity models have been studied by Carroll \emph{et al} \cite{Carroll}.
In this context, many authors have
demonstrated interest by exploring some of their interesting properties and characteristics,
such as: structure formation and stability conditions of neutron stars, which was
studied by \cite{Santos,Tsujikawa} and the gravitational collapse of  spherically
symmetric perfect fluids by \cite{Sharif}.

Apart from the cosmological solutions, gravitational theories must be tested also
in the astrophysical Level to get a more realistic picture of the gravitational fields.
In this sense, strong gravitational regimes found in
relativistic stars could discriminate General Relativity from its generalizations \cite{Psaltis}.
 However, it should be remembered that present day astronomical observational data suggests that neutron stars
 can be used to investigate possible deviations from GR and act as probes for any modified gravity model. These ideas will recent added impetus in the coming decade when the Square Kilometre Array experiment is fully functional.
 At the same time, the structure of compact stars in perturbative f(R) gravity has been studied in
 \cite{Arapoglu}. Another class of interesting models on hydrostatic equilibrium of stellar structure that have been investigated in the framework of f(R)-gravity by \cite{Stabile}.
  The primary motivation for finding  exact interior solutions to Einstein's
 Field equations in f(R) gravity for their static and spherically symmetric case is that they can be
 used to model physically relevant compact objects.
 With the estimate values for masses and radii from different compact objects, the recent
 observational prediction fails to be compatible with the standard neutron star models because
densities of  nuclear matter is much lower than the densities of such compact objects.
This situation can theoretically be explained by the existence of two different kinds of
interior pressures namely, radial and transverse pressure. In this regard, the study of  models of anisotropic
stellar distributions  has received  great interest in GR as well in modified theories of gravity. It is required of such models that elementary  requirements are met in order to be physically viable. For example, it is expected that the energy conditions are satisfied, the interior and exterior spacetimes are smoothly matched across a suitable hypersurface and that causality is not violated.
Having these in mind, the motivation is to study models of the anisotropic compact
stars in the framework of f(R) gravity by admitting non-static conformal symmetry using the
diagonal tetrad method. Indeed the model considered here corresponds to the standard Einstein theory in view of the choice of an action principle that is linear in the Ricci scalar. The consequence is that the equations of motion are the usual second order variety. If the more general Starobinksy model were selected then the equations of motion would involve derivatives of the order of four. This case will be pursued in another study.  At first, Maartens and Maharaj \cite{Maartens} have studied the static spherical
distributions in charged imperfect fluids admitting conformal symmetry and now its become a popular
model to established a natural relationship between geometry and matter source, with the help of the
Einstein field equations. In the same context Das et al. \cite{Das}, obtained
a set of solutions describing the interior of a compact star under f(R, T) gravity and
Ray et al. \cite{Ray} have studied electromagnetic mass models for static
spherical distribution admitting CKV.

The study of anisotropic spherically symmetric distributions have a long history. Such models are heavily under-determined and admit a large number of solutions.  Bowers and Liang \cite{bowers} presented a simple anisotropic model with constant density. The mass implications in anisotropic distributions were revisited by Baracco {\i et al} \cite{barraco} and it was shown that the well known Buchdahl mass limit applicable to isotropic spheres may be violated. Ivanov \cite{ivanov} established mass-radius bounds for anisotropic fluids with spherical symmetry. Mathematical prescriptions such as conformal flatness were studied by Stewart \cite{sterwart} who postulated forms for the mass function to generate exact models.  Nonstatic spheres admitting a one parameter group of conformal motions were studied by Herrera and Ponce de Leon \cite{herrera} but without specifying a linear equation of state.

Motivated by the above discussion, we study the compact stars solutions with their interesting
properties and characteristics in the context of f(R) theory of gravity admitting initially
non-static conformal symmetry and with a specified equation of state. The paper is organized as follows:
in Section II, we review f(R) modified theories of gravity whereas in Section III,
the CKVs are formulated. In Section IV, we formulate the system of gravitational field equations, then in Section V we describe the star interior, using the spherically symmetric line element and CKVs.
In Section VI, we have discussed various physical features of the model such as energy
conditions, equilibrium condition by using Tolman-Oppenheimer-Volkoff (TOV)
equation and stability issue. Finally in Section VII, we conclude with a summary and discussions.

\section{Basic Mathematical Formulation in f(R) Gravity}

Let us consider the following action in f(R) gravity of the form
\begin{equation}\label{1}
\pounds =\int dx^{4}\sqrt{-g}[f(R)+{L_{m}}],
\end{equation}
where $g$ is the determinant of the space-time metric $g_{\mu\nu}$ assuming that,
$ 8 \pi G=1$ and $L_{m}$ is the matter Lagrangian. Here f(R) is a
general function of the Ricci scalar, R. Within the context of the
metric formalism, variation of the action Equation (1) with respect to $g_{\mu\nu}$
generates the field equations

\begin{equation}\label{2}
G_{\mu\nu}= R_{\mu\nu}- \frac{1}{2}Rg_{\mu\nu}=T_{\mu\nu}^{(curv)}+ T_{\mu\nu}^{(matter)},
\end{equation}
where the term $T_{\mu\nu}^{(matter)}$ represents the energy momentum tensor of the matter scaled by a factor of
$1 /{f'(R)}$ and $T_{\mu\nu}^{(curv)}$ is an extra stress-energy tensor contribution by the
f(R) modified gravity theory and is given by

\begin{eqnarray}
T_{\mu\nu}^{(curv)}&= &\frac{1}{F(R)}
\left[~\frac{1}{2}g_{\mu\nu}\left(f(R)-RF(R)\right)\right.\nonumber\\
&\;&\left.+F(R)^{;\alpha\beta}
\left(g_{\mu\alpha}g_{\nu\beta}-g_{\mu\nu}g_{\alpha\beta}\right)\right].\nonumber\
\end{eqnarray}

In the above equation, $F(R) \equiv df/dR$ and primes denote derivatives with respect to Ricci scalar $R$.
Now, considering a static spherically symmetric distribution of matter, the  line element may be expressed in the form
\begin{equation}\label{1}
ds^{2}=-e^{\nu(r)}dt^{2}+e^{\lambda(r)}dr^{2}+r^{2}\left(d\theta^{2}+\sin^{2}\theta d\phi^{2}\right),
\end{equation}
where the metric functions $\nu(r)$ and $\lambda(r)$ are functions of the radial coordinate which may be utilised to compute the
the surface gravitational redshift and the mass function, respectively. From the metric it is possible to
construct asymptotically flat spacetimes when $ \nu(r) \rightarrow 0$ and $ \lambda(r) \rightarrow 0$
as $ r \rightarrow \infty$.

To obtain some particular strange star models we assume that the material composition
filling the interior of the compact object is anisotropic in nature
and accordingly we express the energy-momentum tensor in the form
\begin{equation}
T_{\alpha\beta}^{m}=\left(\rho + p_{t}\right)u_{\alpha}u_{\beta}-p_{t}g_{\alpha\beta}+\left(p_{r}-p_{t}\right)v_{\alpha}v_{\beta},
\end{equation}
where $\rho$ depicts energy density, $p_{r}$ is the radial pressure, $p_{t}$ to be the tangential pressure, respectively. In the expression $u_{\alpha}$ is the four-velocity of the fluid and $v_{\alpha}$
corresponds to the radial four vector. Note that with the assumption  of an anisotropic particle pressure, the field equations may be solved by an infinite number of metrics. Additional restrictions based on physical grounds must be introduced to model realistic objects. Customarily an equation of state relating the pressure and the density may be imposed, however, this approach runs into sever mathematical problems making the equation of pressure isotropy impractical to solve. In this work, we elect to utilise a geometric prescription. New classes of anisotropic star solutions admitting conformal motions are sought and investigated.

\section{Conformal Killing Vector}

To establish a natural relationship between geometry and matter through the Einstein's GR,
and searching for an exact solutions by convincing approach we use a non-static conformal
symmetry. Following the idea a work has recently been considered in \cite{das}.
For instance, one may specify the vector $\xi$ as the generator of  this conformal symmetry,
and the metric $g$ is conformally mapped onto itself along $\xi$ so that  the following relationship
\begin{equation}
\mathcal{L}_{\xi}g_{ij}= \psi g_{ij},
\end{equation}
is obeyed. Here $\mathcal{L}$ represents the Lie derivative operator of the metric tensor
whereas $\psi$ is the conformal \textbf{killing} vector.

As for the choice of  the vector $\xi$, generates the conformal symmetry within the framework of the standard GR, then the metric $g$ is conformally mapped onto itself along $\xi$.
According to B$\ddot{o}$hmer et al. \cite{Bohmer}, for the choice of static metric,
neither $\xi$ nor $\psi$ needs be static. It should be noted that if $\psi=0$ then
Eq. (6) gives the Killing vector, if $\psi=constant$ then it yields a homothetic
vector and if $\psi=\psi(x,t)$ then it gives conformal vectors.
Furthermore, the spacetime becomes asymptotically flat when $\psi=0$, which also implies that the Weyl tensor will vanish. All such conformally flat spacetimes have been found. They are either the
Schwarzschild interior solution in the case of no expansion or the Stephani \cite{stephani} stars if expansion is occurring. Accumulated all of this properties of CKV have been effectively provide a deeper insight
Accumulated all of this properties of CKV have effectively provided a comprehensive picture
of the underlying spacetime geometry. Here Eq. $(5)$ implies that
\begin{equation}
\mathcal{L}_{\xi}g_{ik}= \xi_{i;k}+ \xi_{k;i}=\psi g_{ik},
\end{equation}
with $ \xi_{i}=g_{ik}\xi^{k}$. From Eq. $(4)$ and $(7)$, the following expressions \cite{Bohmer} are obtained as

\[\xi^{1}\nu'=\psi, ~\xi^{4}= \text{const.}, ~\xi^{1}= \frac{\psi r}{2} ~~\text{and} ~~\xi^{1}\lambda' + 2\xi_{,1}^{1}=\psi,\]
where $1$ and $4$ stand for spatial and temporal coordinates $r$ and $t$ respectively.
Then Eq. (6) provides the following relationship by using the metric (2), we have
\begin{equation}
e^{\nu}=C_{0}^{2}r^{2},
\end{equation}
\begin{equation}
e^{\lambda}=\left[\frac{C}{\psi}\right]^{2},
\end{equation}
\begin{equation}
\xi^{i}=C_{1}\delta_{4}^{i}+ \left[\frac{\psi r}{2}\right]\delta_{1}^{i},
\end{equation}
where $C$, $C_{0}$ and $C_{1}$ are all integration constants.

\section{The Field equations }

The tetrad matrix for the metric $(4)$ is defined as
\begin{equation}
e_{\mu}^{i}= diag\left[\sqrt{C_{0}^{2}r^{2}},e^{\frac{\lambda}{2}},r,r \sin{\theta}\right].
\end{equation}
Therefore, the Ricci scalar is determined as
\begin{equation}R(r)= -\frac{2}{r^{2}}+ e^{-\lambda(r)}\left(\frac{6}{r^{2}}-\frac{3\lambda'(r)}{r}\right).
\end{equation}
Using the space time metric given by Eq. (4),
the modified field Eq. (2) generates the equations
\begin{eqnarray}
\rho &= & -e^{-\lambda}F'' +e^{-\lambda}\left(\frac{\lambda'}{2}-\frac{2}{r}\right)F'
+\frac{e^{-\lambda}}{r^{2}}\left(\frac{\nu'' r^{2}}{2}\right.\nonumber\\
&\;&\left.+\frac{\nu'^{2}r^{2}}{4}-\frac{\nu'\lambda' r^{2}}{4}+\nu'r\right)F-\frac{1}{2}f,
\end{eqnarray}

\begin{eqnarray}
p_{r} &= & e^{-\lambda}\left(\frac{\nu'}{2}+\frac{2}{r}\right)F'-\frac{e^{-\lambda}}{r^{2}}\left(\frac{\nu'' r^{2}}{2}+\frac{\nu'^{2}r^{2}}{4}\right.\nonumber\\
&\;&\left.-\frac{\nu'\lambda' r^{2}}{4}-\lambda'r\right)F+\frac{1}{2}f,
\end{eqnarray}

\begin{eqnarray}
p_{t} &= & -e^{-\lambda}F''+e^{-\lambda}\left(\frac{\nu'}{2}-\frac{\lambda'}{2}+\frac{1}{r}\right)F'
-\frac{e^{-\lambda}}{r^{2}}\left(\frac{\nu'r}{2}\right.\nonumber\\
&\;&\left.-\frac{\lambda'r}{2}-e^{\lambda}+1\right)F+\frac{1}{2}f,
\end{eqnarray}
governing the evolution of the fluid.
In the following we focus on several physically relevant f(R) gravity models and
seek interior solutions of the Einstein field equations with conformal Killing vector,
that can model compact objects. The field equations of f(R) modified theories
are highly non-trivial to solve, so the entire calculations are highly
dependent on the equation ansatz. We choose to solve the continuity equation for decoding the
system of equations for our purpose.

\section{Solutions:}

As a first step in our analysis we  consider solutions for
f(R)=R and f(R) = aR + b, where $a$ and $b$ are purely constant quantities. Note that this prescriptions is essentially the general relativity case without and with a cosmological constant.
We further assume that the standard matter for a fluid distribution obeys the
equation of state
\begin{equation}
p_{r}=\omega\rho
\end{equation}
where the state parameter $w$ is a constant and , $ 0< w < 1 $  for causality. That is the speed of sound remains subluminal provided that $0 \leq \frac{dp}{d\rho} < 1$. Now we determine  the
solutions under the above conditions.

\subsection{Case I: $f(R) = R$ }
4.1.1: When {$p_{r} = p_{t} = p$}

Indeed, by using Eqs. $(10 - 14)$, we obtain the solution of this ordinary
differential equation in the form
\begin{equation}
\rho(r)=\frac{1}{r^{2}}+\frac{3B}{2},
\end{equation}
\begin{equation}
p(r)=\frac{1}{2r^{2}}-\frac{3B}{2},
\end{equation}
while the conformal factor, Ricci scalar and mass functions are given by
\begin{equation}
\psi=C\left[\frac{1}{2}-\frac{B}{2}r^{2}\right]^{\frac{1}{2}},
\end{equation}
\begin{equation}
R(r)=-6B+\frac{1}{r^{2}},
\end{equation}
\begin{equation}
e^{-\lambda(r)}=\frac{1}{2}-\frac{B}{2}r^{2},
\end{equation}
where $B$ is an integration constant. In this case, the velocity of
sound  $v_{s}^{2}$ is given by
\begin{equation}
\frac{dp}{d\rho}=\frac{1}{2}.
\end{equation}
On account of the assumption of isotropic particle pressure, we have not invoked the relationship $p=\omega \rho$ so as not to over-determine the system. However, the solution above does indeed exhibit an equation of state of the form $p = p(\rho)$ consistent with perfect fluid distributions.  Observe that setting $B=0$ gives the Saslaw et al.  \cite{saslaw} isothermal fluid model with both  the energy density and pressure obeying the inverse square law and obviously the barotropic linear equation of state $p = \omega \rho$. That the metric potential $\lambda$ is not constant in this model may be attributed to the presence of the conformal symmetry.
\begin{figure*}[thbp]
\begin{center}
\begin{tabular}{rl}
\includegraphics[width=7 cm]{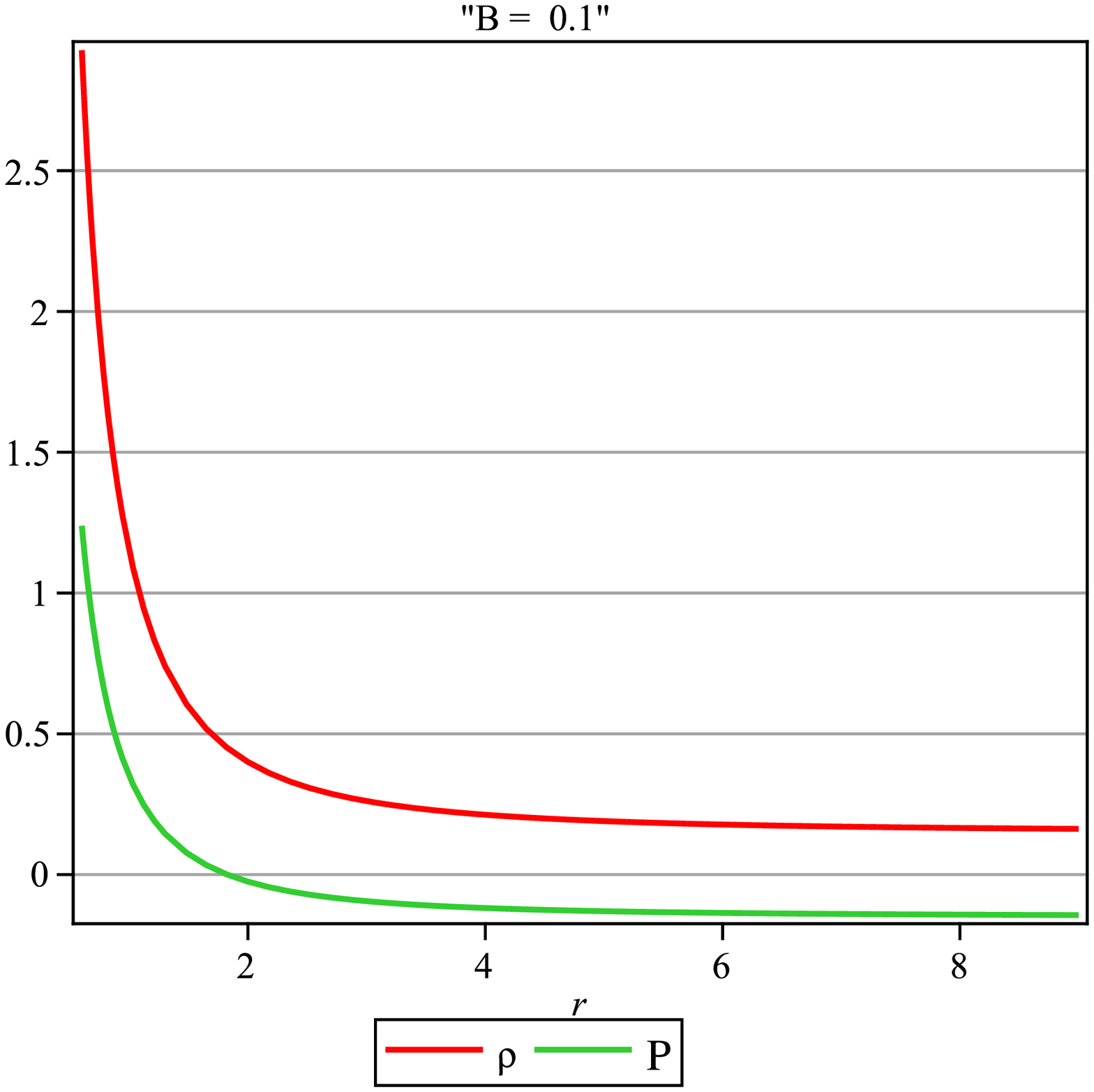}&
\includegraphics[width=7 cm]{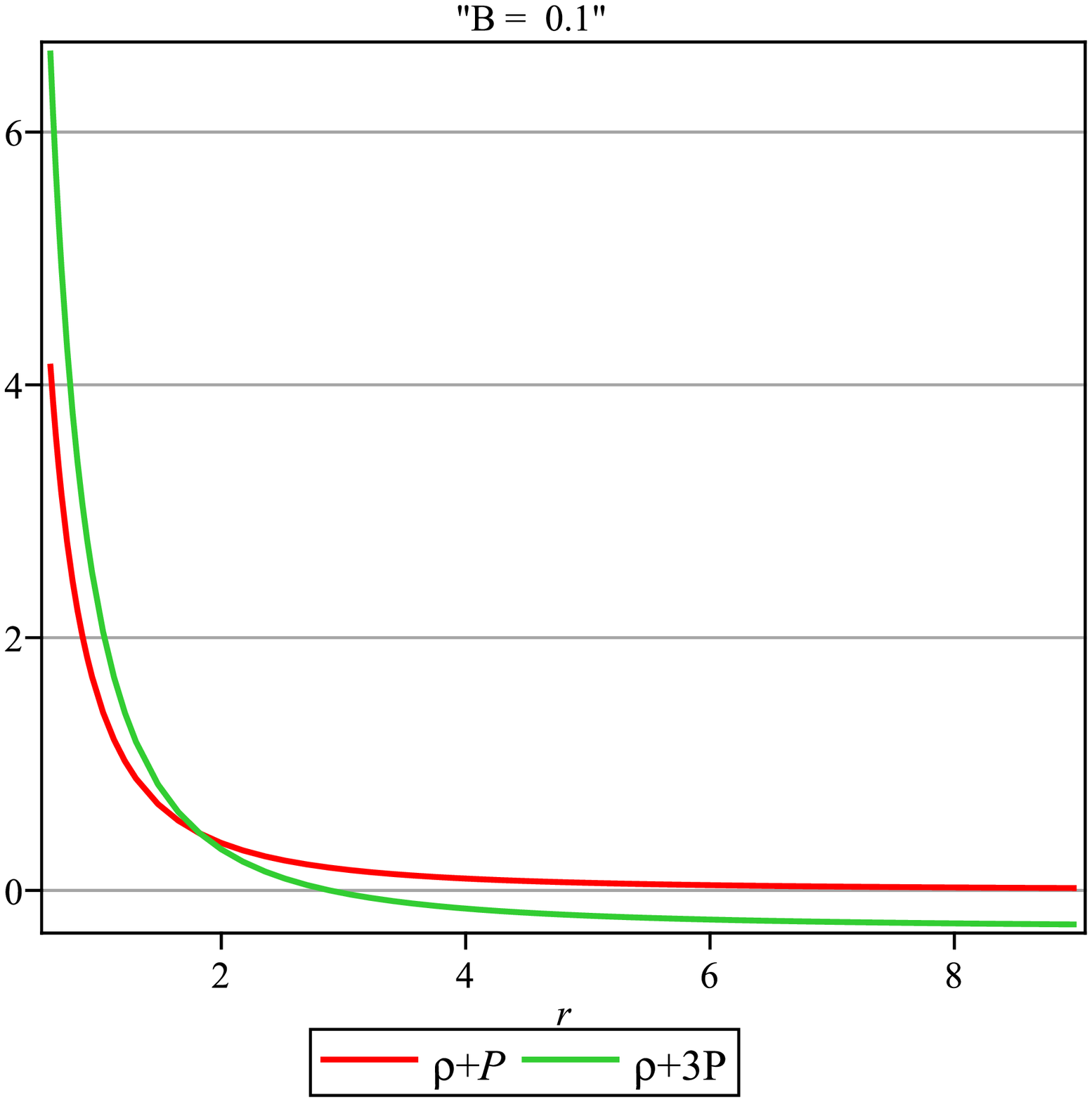}\\
\end{tabular}
\end{center}
\caption{Variation of density and pressure with Energy conditions of the first model:
We have plotted the energy density and pressures, respectively in the
left panel whereas in the right panel we have drawn the behavior of different energy conditions
when the constant term 'B = 0.8' assuming a positive value. Here we see that null energy
conditions (NEC) is satisfied but strong energy conditions (SEC) is violated
for isotropic case.}
\end{figure*}

\newpage
4.1.2: When $p_{r}\neq p_{t}$

Using the Eqs. $(10 - 15)$, we obtain the solution when matter distribution
is anisotropic in nature and it given by
\begin{equation}
\rho(r)=-\frac{6D}{\omega}r^{\frac{-(6+\omega)}{(3+\omega)}}+\frac{2}{(3+\omega)}r^{-2},
\end{equation}
\begin{equation}
p_{r}= -6Dr^\frac{-(6+\omega)}{(3+\omega)}+\frac{2\omega}{(3+\omega)}r^{-2},
\end{equation}
and
\begin{equation}
p_{t}=\frac{-\omega^{2}-8\omega-3}{2\omega(3+\omega)}r^{-2}-\frac{6D}{\omega}r^{\frac{-(6+\omega)}{(3+\omega)}} +3Dr^{-3}.
\end{equation}
and the other set of expressions are given by
\begin{equation}
\psi=C\left[\frac{1+\omega}{3+\omega}+Dr^{\frac{\omega}{3+\omega}}\right]^{\frac{1}{2}},
\end{equation}
\begin{equation}
e^{-\lambda(r)}=\frac{1+\omega}{3+\omega}+Dr^{\frac{\omega}{3+\omega}},
\end{equation}
\begin{equation}
R(r)=-\frac{2}{r^{2}}+\frac{6(1+\omega)}{(3+\omega)r^{2}}+6Dr^{-3}+\frac{3(3+\omega)D}
{\omega}r^{\frac{-(6+\omega)}{(3+\omega)}}.
\end{equation}
where $D$ is integration constant. The sound velocity $v_{s}^{2}$ is determined as
\begin{equation}
\frac{dp_{t}}{d\rho}=\frac{\frac{\omega^{2}+8\omega+3}{2\omega(3+\omega)r^{3}}
+\frac{6D(6+\omega)}{\omega(3+\omega)}r^{\frac{-2\omega-9}{3+\omega}}-9Dr^{-4}}{\frac{6D(6+\omega)}
{\omega(3+\omega)}r^{\frac{-2\omega-9}{3+\omega}}-\frac{4\omega}{(3+\omega)r^{3}}}.
\end{equation}
Note that in view of the latitude afforded by the assumption of pressure anisotropy the field equations were augmented by the equation of state $p = \omega \rho$.

\begin{figure*}[thbp]
\begin{center}
\begin{tabular}{rl}
\includegraphics[width=6 cm]{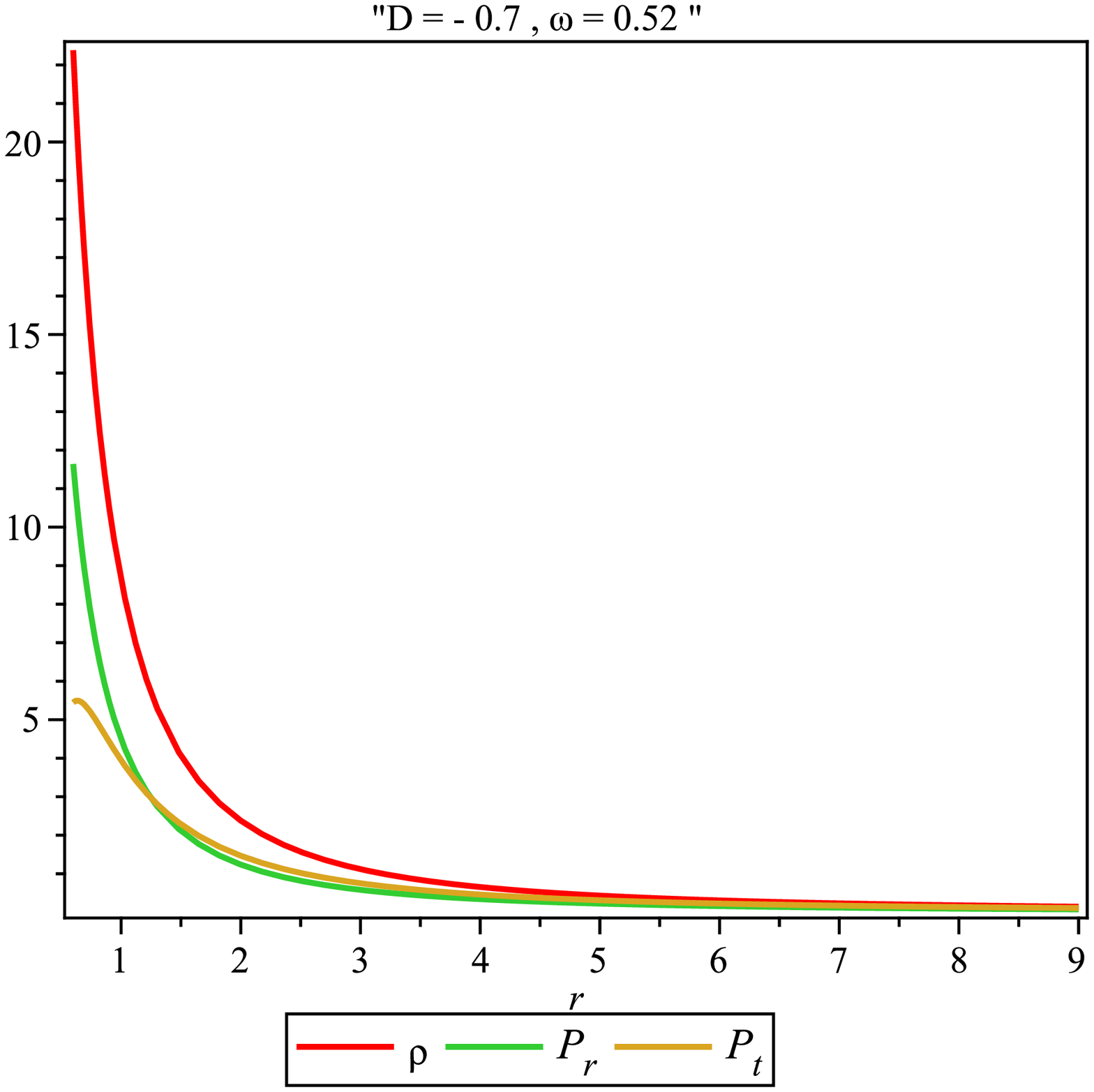}&
\includegraphics[width=6 cm]{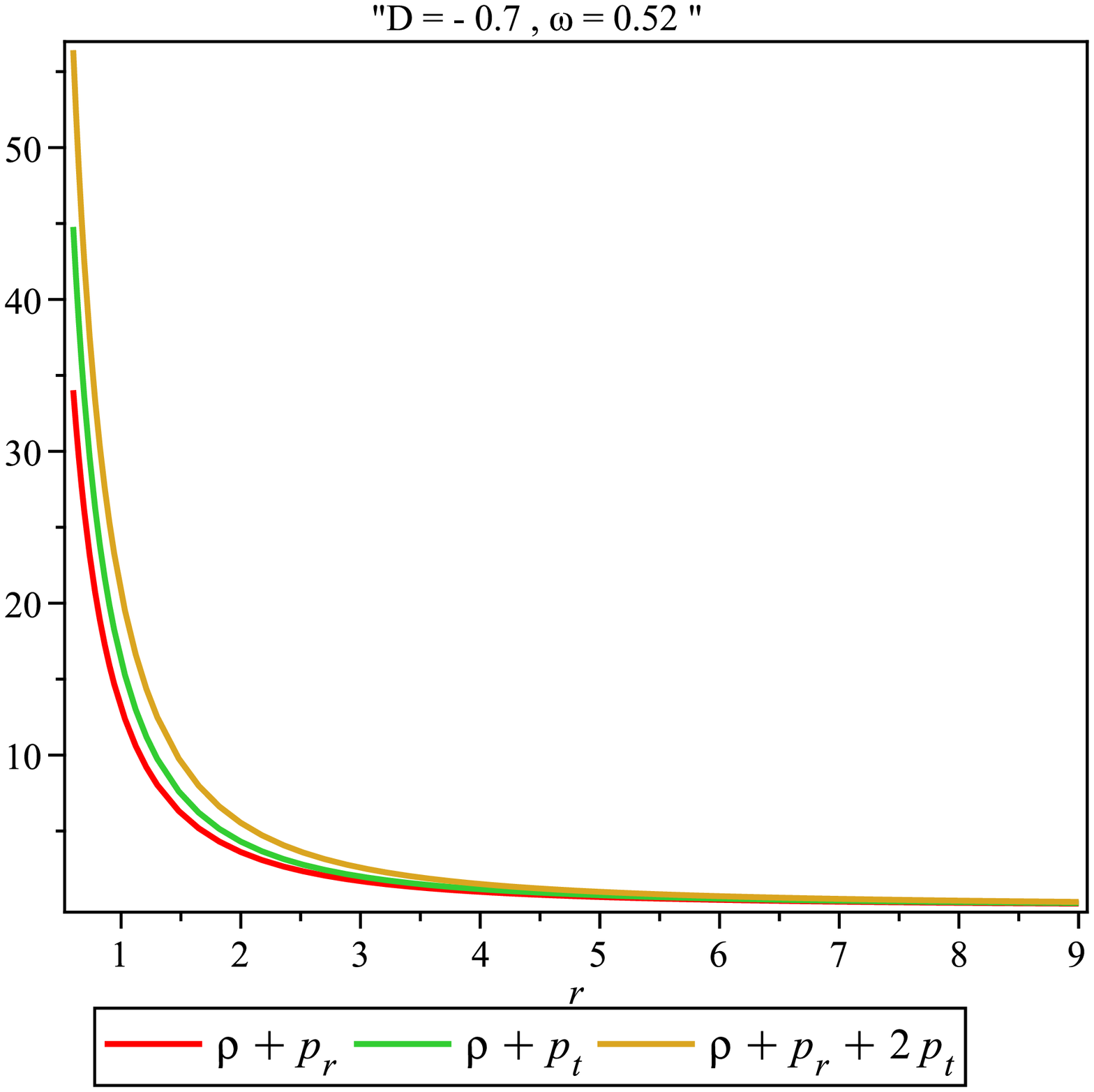}\\
\end{tabular}
\end{center}
\caption{Variation of density and pressure with Energy conditions of the second model:
In the left panel the functions of energy density and pressures have been plotted
with the radial coordinate and in the right panel we have plotted all the energy conditions
inside the star when f(R) = R and $p_{r}\neq p_{t}$. The parametric values have been taken for the graphs
are shown in the legend.}
\end{figure*}
\begin{figure*}[thbp]
\begin{center}
\begin{tabular}{rl}
\includegraphics[width=6 cm]{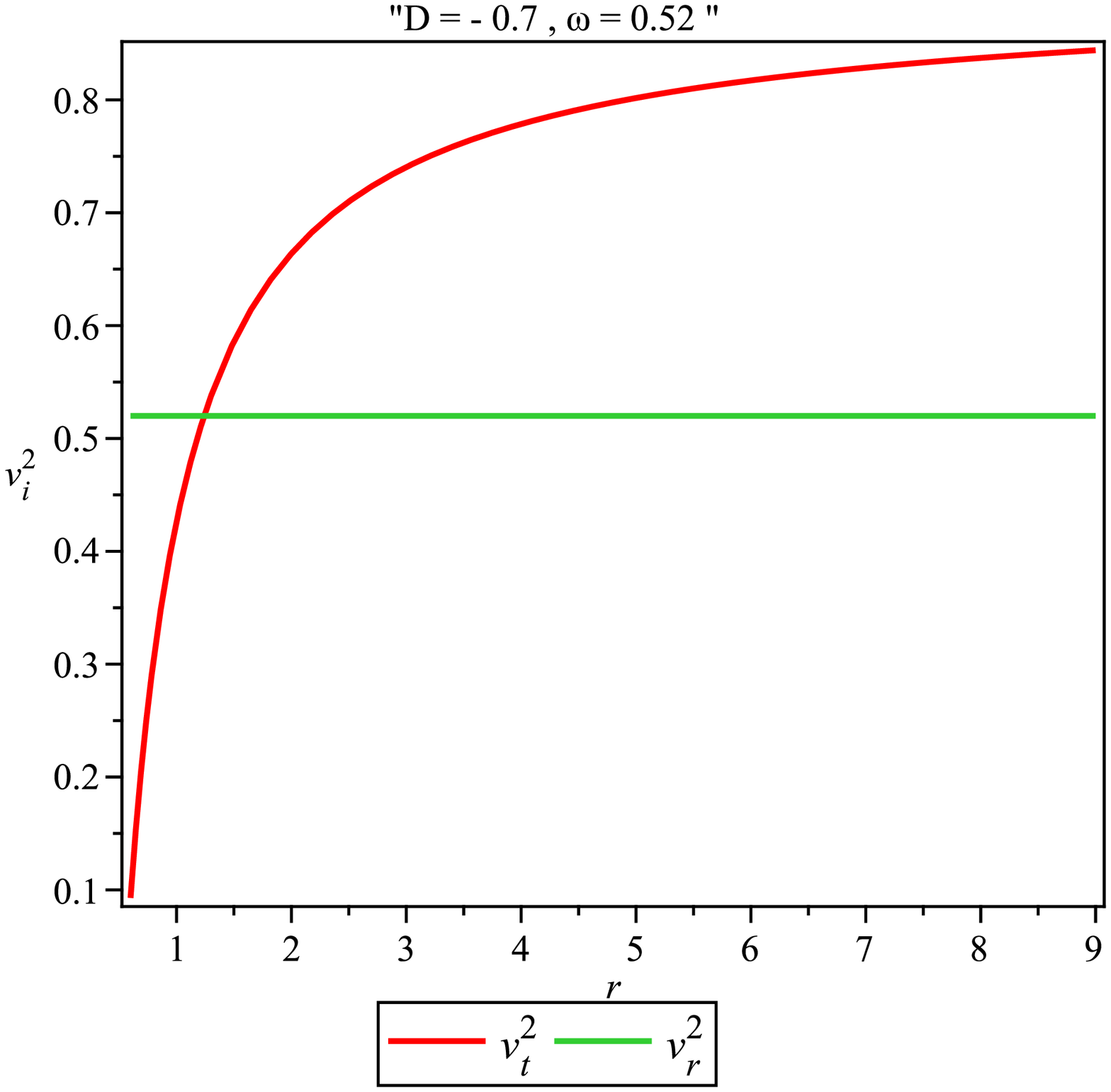}&
\includegraphics[width=6 cm]{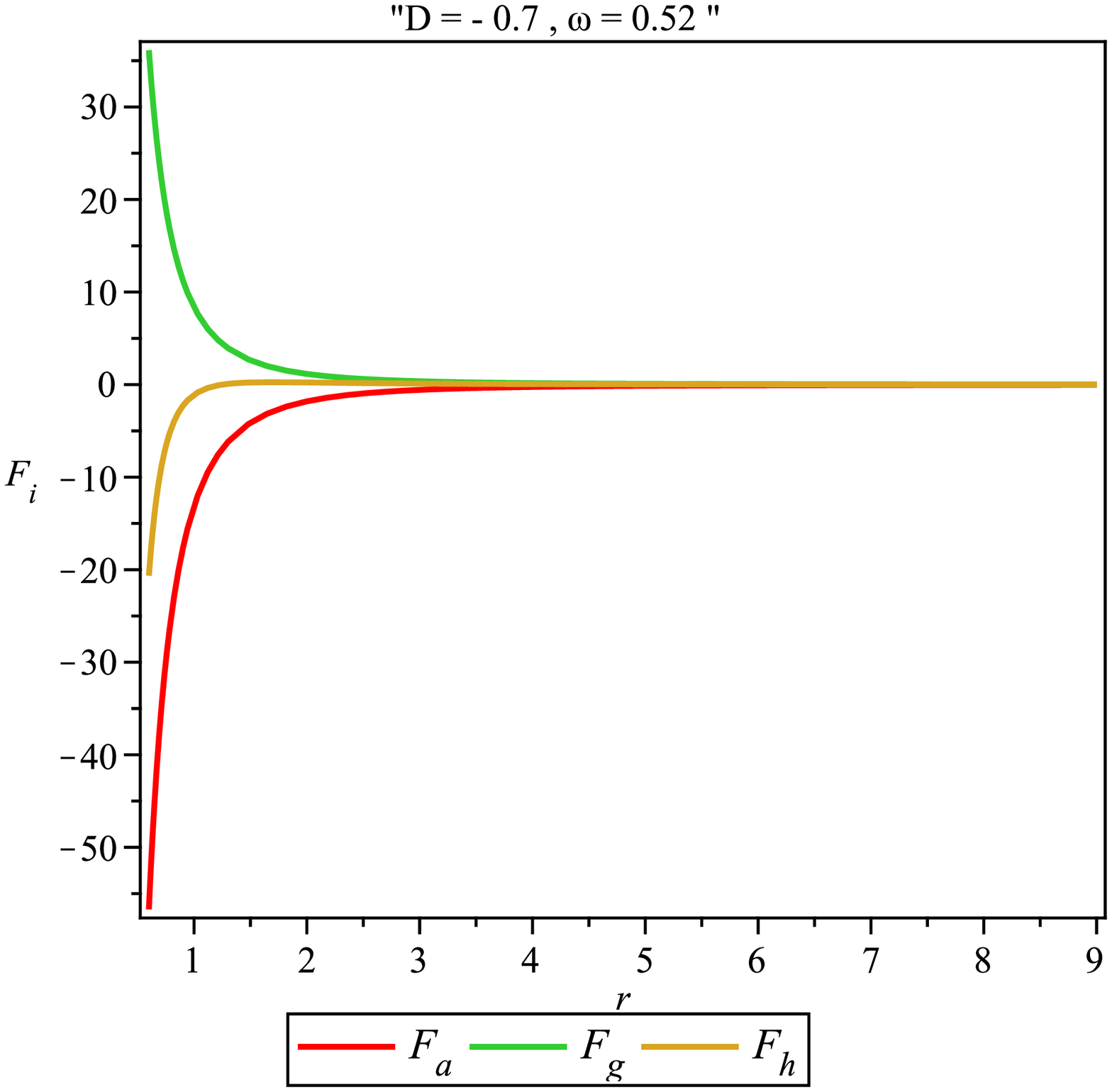}\\
\end{tabular}
\end{center}
\caption{Illustrative plot of square of sound speed (in the left panel), whereas we have plotted
variation of different forces in the right panel so that the system is subjected to the equilibrium
position for the f(R) = R solution. The parameter values have been taken for the graphs
are shown in the legend.}
\end{figure*}

\subsection{Case II: f(R)= aR + b}

Now we are going to consider the second case where the effects of the cosmological constant enter the picture through the constant $b$. The initial case is that of

4.2.1: When {$p_{r}=p_{t}=p$}

Using Eqs. $(10 - 14)$ and assuming the isotropic condition,
provides the following general solution
\begin{equation}
\rho(r)=\frac{a}{2r^{2}}-3aE-\frac{ab}{2},
\end{equation}
\begin{equation}
p(r)=3aE +\frac{a}{2r^{2}}+\frac{ab}{2},
\end{equation}
and the conformal factor, Ricci scalar and mass functions are given by
\begin{equation}
\psi=C\left[\frac{1}{2}+Er^{2}\right]^{\frac{1}{2}},
\end{equation}
\begin{equation}
R(r)=12E+\frac{1}{r^{2}},
\end{equation}
\begin{equation}
e^{-\lambda(r)}=\frac{1}{2}+Er^{2},
\end{equation}
where $E$ is an integration constant. The sound velocity $v_{s}^{2}$ is then obtained as
\begin{equation}
\frac{dp}{d\rho}=1.
\end{equation}
which is the extreme value for the sound speed ordinarily associated with a stiff fluid.
\begin{figure*}[thbp]
\begin{center}
\begin{tabular}{rl}
\includegraphics[width=6 cm]{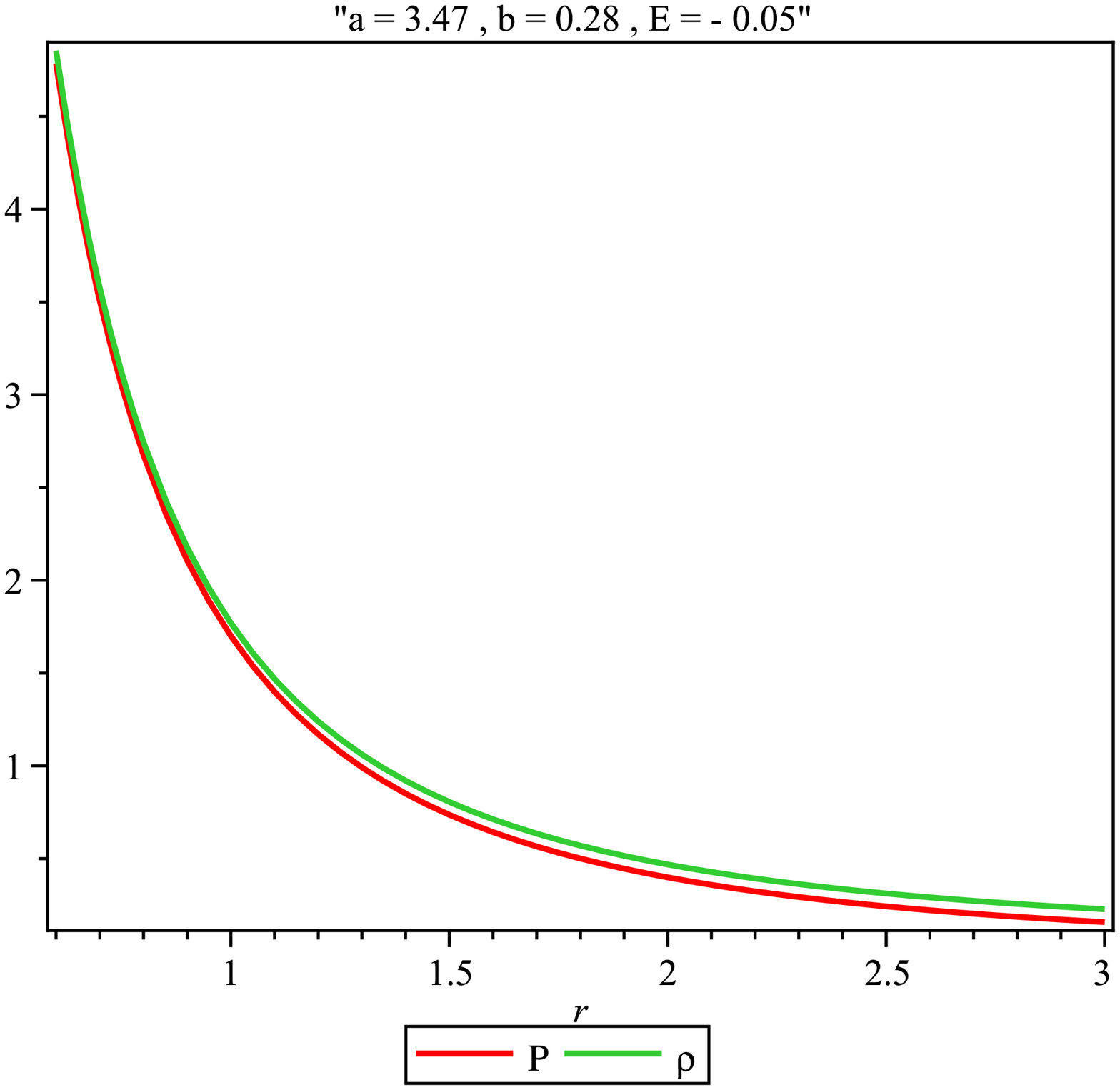}&
\includegraphics[width=6 cm]{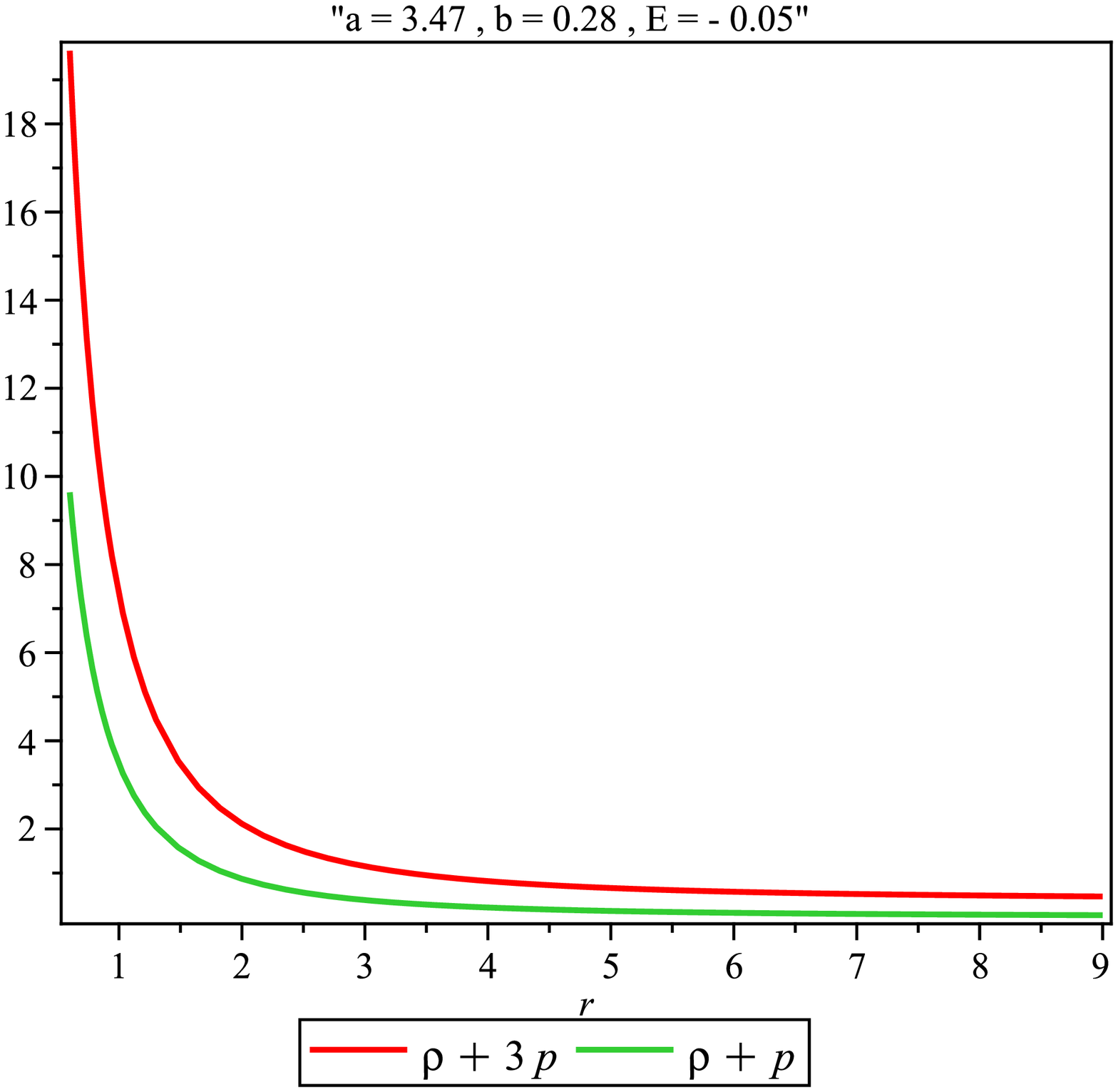}\\
\end{tabular}
\end{center}
\caption{Plots for variation of density and pressure with Energy conditions when f(R)= aR + b:
In the left panel the functions of energy density and pressures have been plotted
with the radial coordinate and in the right panel we have plotted all the energy conditions
inside the star for isotropic model of the fluid sphere. The parameter values have been taken for the graphs
are shown in the legend.}
\end{figure*}
\begin{figure*}[thbp]
\begin{center}
\begin{tabular}{rl}
\includegraphics[width=6 cm]{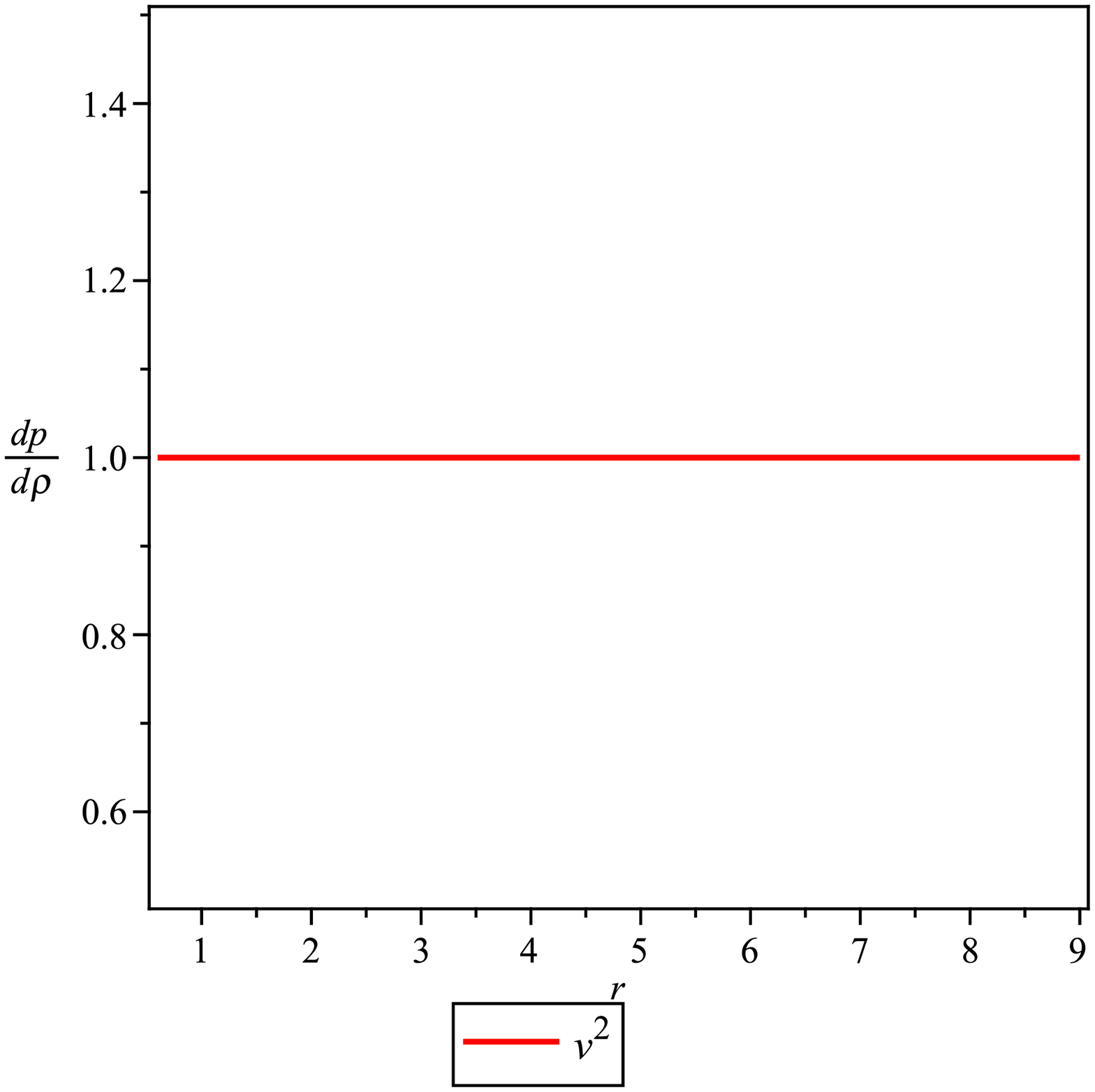}&
\includegraphics[width=6 cm]{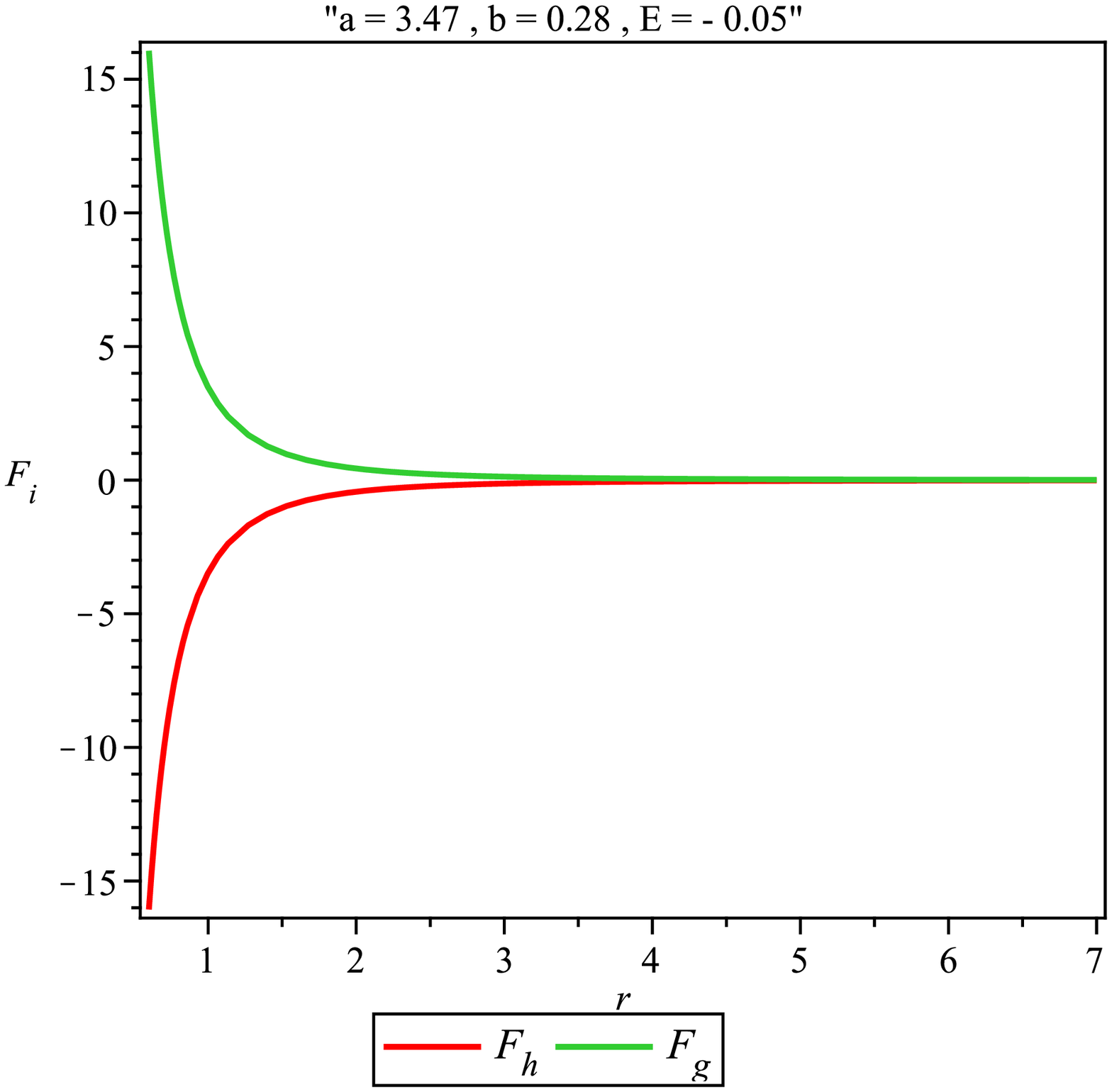}\\
\end{tabular}
\end{center}
\caption{Illustrative plot of square of sound speed (in the left panel), whereas we have plotted
variation of different forces in the right panel, so that the system is subjected to the equilibrium
position for the f(R) = aR+b solution. The parameter values have been taken for the graphs
are shown in the legend.}
\end{figure*}

4.2.2: When $p_{r}\neq p_{t}$

Using Eqs. $(10 - 14)$, and Eq. $(15)$ the differential equations when solved give
\begin{equation}
\rho(r)=\frac{b(1+\omega)}{\omega}+\frac{2a}{(3+\omega)r^{2}}+\frac{3aH}{\omega}r^{\frac{-3(1+\omega)}{\omega}},
\end{equation}
\begin{equation}
p_{r}= b(1+\omega)+\frac{2a\omega}{(3+\omega)r^{2}}+3aHr^{\frac{-3(1+\omega)}{\omega}},
\end{equation}
\begin{equation}
p_{t}= \frac{a(1+\omega)}{(3+\omega)r^{2}}-\frac{b}{2\omega}+H(2a-\frac{3}{\omega})r^{\frac{-3(1+\omega)}{\omega}},
\end{equation}
with other functions are given by
\begin{equation}
\psi=C\left[\frac{1+\omega}{3+\omega}-\frac{br^{2}}{6a}+Hr^{\frac{-(3+\omega)}{\omega}}\right]^{\frac{1}{2}},
\end{equation}
\begin{equation}
R(r)=\frac{4\omega}{(3+\omega)r^{2}}-\frac{b(1+2\omega)}{a\omega}+\frac{3\omega-9}{\omega}Hr^{\frac{-3(1+\omega)}{\omega}},
\end{equation}
and
\begin{equation}
e^{-\lambda(r)}=\frac{1+\omega}{3+\omega}-\frac{br^{2}}{6a}+Hr^{\frac{-(3+\omega)}{\omega}},
\end{equation}

The sound velocity $v_{s}^{2}$ can be found out as
\begin{equation}
\frac{dp_{t}}{d\rho}= \frac{\frac{2a(1+\omega)}{(3+\omega)r^{3}}+\frac{3(1+\omega)}{\omega}(2a-\frac{3}{\omega})
r^{\frac{-3-4\omega}{\omega}}}{\frac{4a}{(3+\omega)r^{3}}+\frac{9aH(1+\omega)}{\omega}r^{\frac{-3-4\omega}{\omega}}}.
\end{equation}
where $H$ is an integration constant.
\begin{figure*}[thbp]
\begin{center}
\begin{tabular}{rl}
\includegraphics[width=6.5 cm]{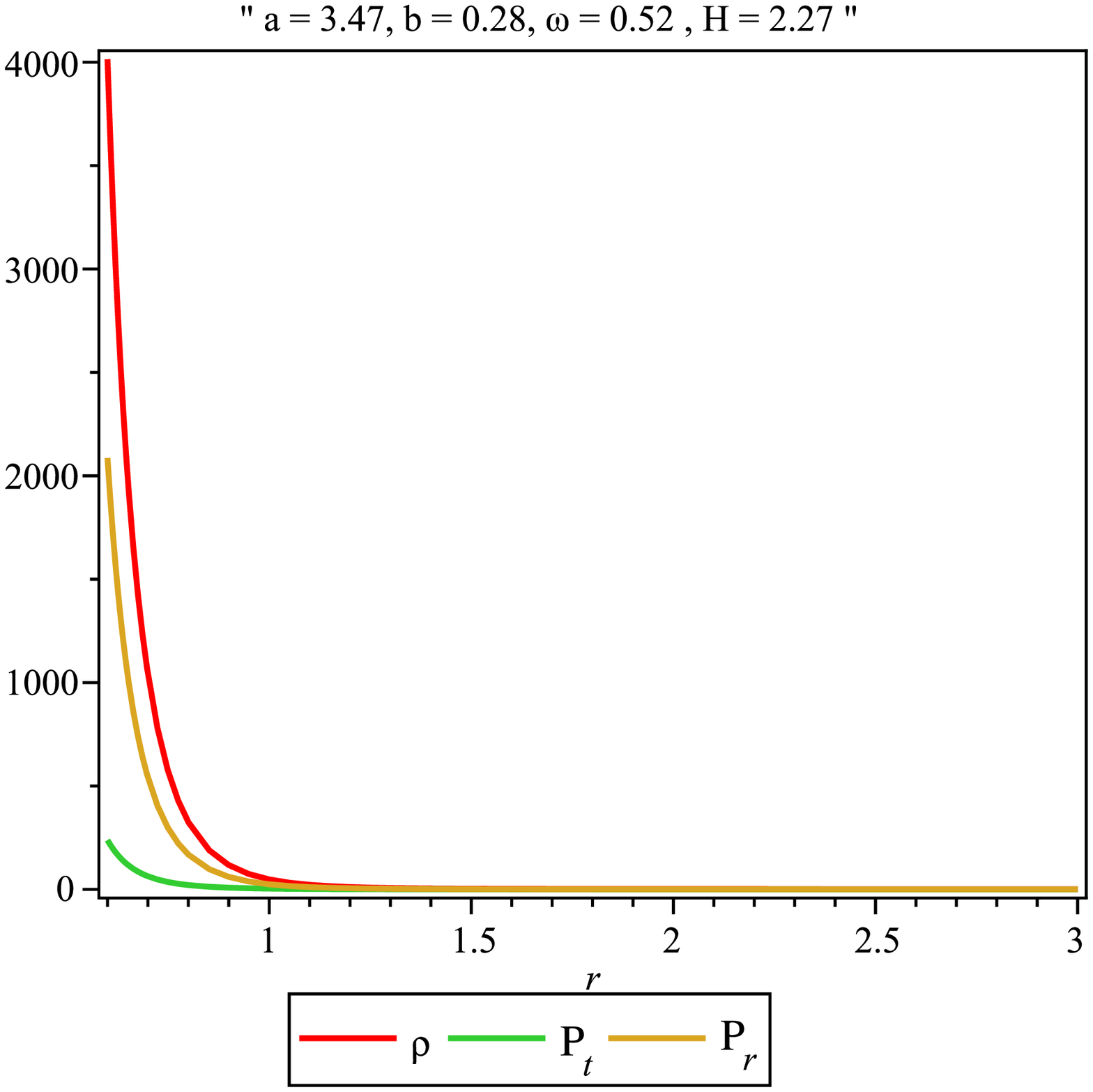}&
\includegraphics[width=6.5 cm]{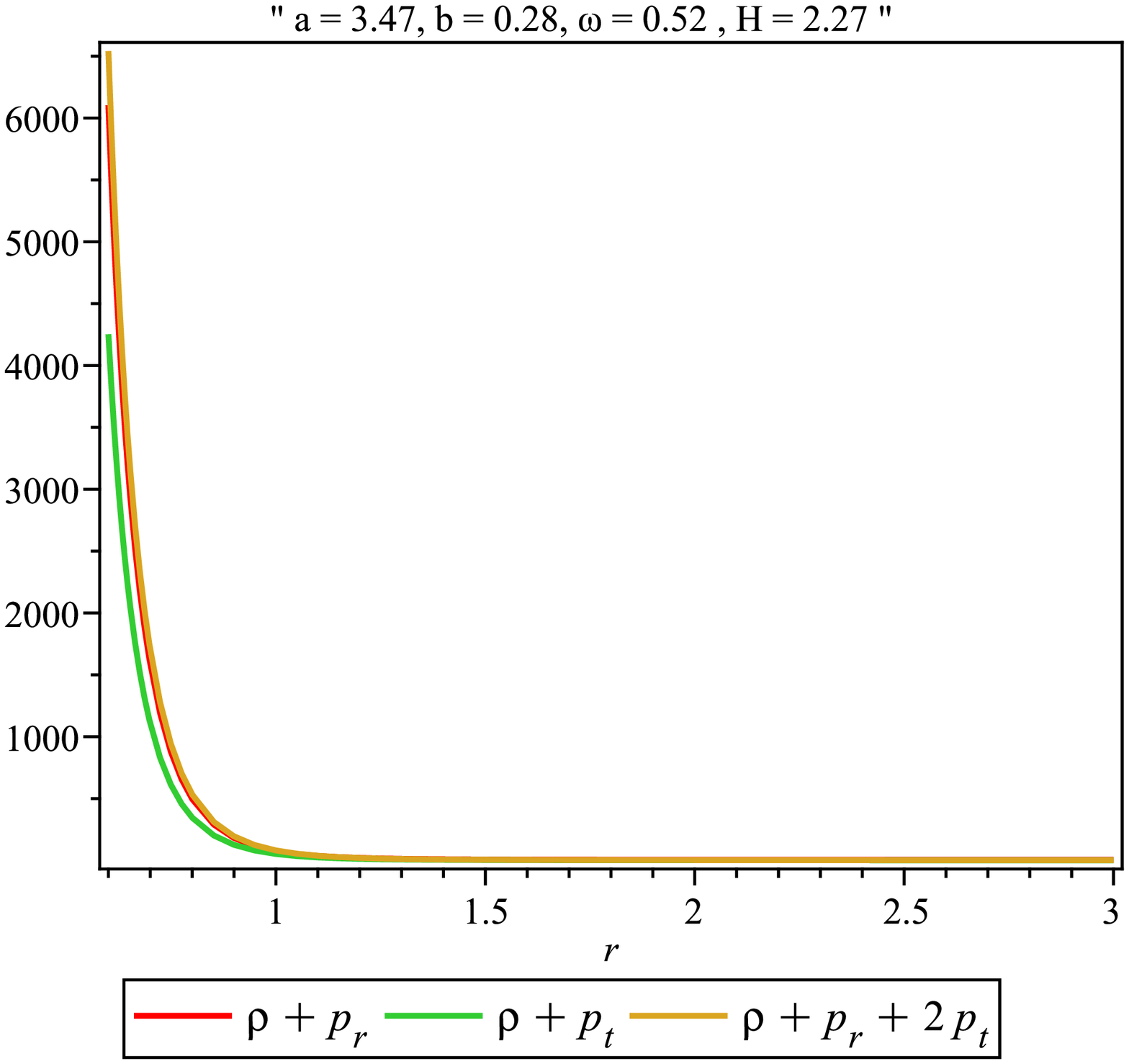}\\
\end{tabular}
\end{center}
\caption{Radial variation of energy density and pressures have been plotted in the left panel,
whereas all the energy conditions inside the star for anisotropic model of the
fluid sphere when f(R) = aR+b solution.}
\end{figure*}
\begin{figure*}[thbp]
\begin{center}
\begin{tabular}{rl}
\includegraphics[width=6.5 cm]{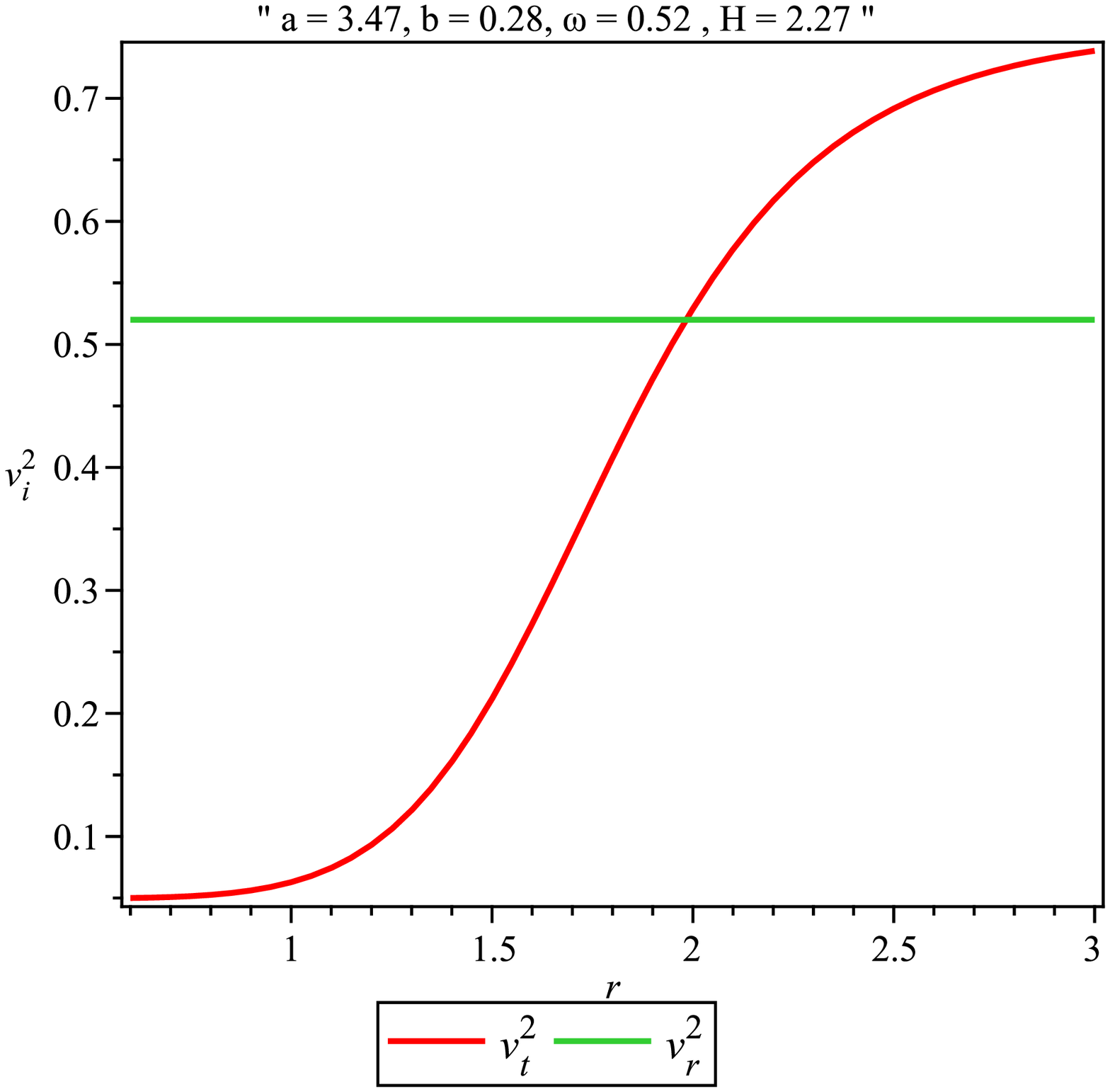}&
\includegraphics[width=6.5 cm]{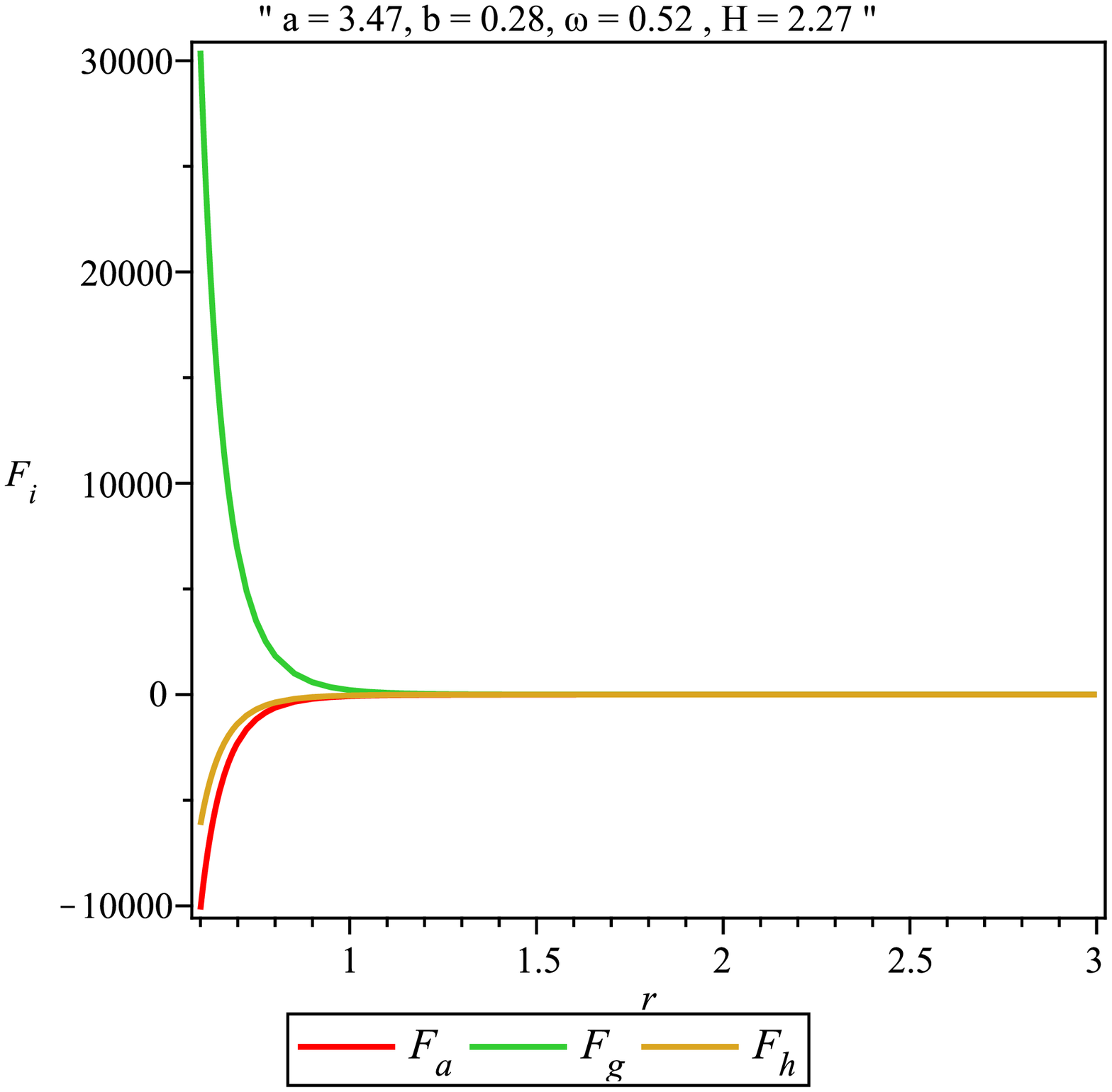}\\
\end{tabular}
\end{center}
\caption{In the left panel the variation of the radial and transverse sound speed have
been plotted, whereas in the right panel the equilibrium condition concerning TOV equation
have been considered for anisotropic f(R) = aR+b solution. The parameter values have been taken for the graphs
are shown in the legend.}
\end{figure*}

\section{Physical features of the model}

\subsection{Energy conditions}
For physically acceptable model, we are going to verify whether the anisotropic
fluid sphere satisfies all the energy conditions or not, namely:
(i) Null energy condition (NEC), (ii) Weak energy
condition (WEC) and (iii) Strong energy condition (SEC) at all points in the interior the star.
We therefore attempt to write down the following inequalities as follows:
\begin{eqnarray}
\textbf{NEC : } \rho(r)+p_r \geq  0, ~\text{and} ~\rho(r)+p_r \geq  0,\\
\textbf{WEC : } \rho \geq  0, \rho(r)+p_r \geq  0, ~\text{and} ~\rho(r)+p_r \geq  0,\\
\textbf{SEC : } \rho(r)+p_r \geq  0, ~\text{and}~ \rho(r)+p_r+ 2p_t \geq  0,
\end{eqnarray}
In figs. 1, 2, 4 and 6 we plot variation of all the energy conditions
inside the compact objects for a given $w = 0.52$ with different values of
constant parameters. In the case of isotropic pressure when f(R) = R, plotted in Fig. 2,
it is evident that null energy conditions (NEC) is satisfied but
strong energy conditions (SEC) is violated for our parametric choice however
for other cases the energy conditions are valid inside the compact star.

\subsection{TOV Equation}
The success of this model lies in its stability under the different
forces namely gravitational, hydrostatic and anisotropic forces.
To examine these we consider the generalized Tolman-Oppenheimer-Volkov (TOV)
equation for anisotropic fluid distribution is given by \cite{leon}

\begin{equation}
-\frac{M_{G}(r)(\rho +p_{r})}{r^{2}}e^{\frac{\lambda-\nu}{2}}-\frac{dp_{r}}{dr}+\frac{2}{r}(p_{t}-p_{r})=0,
\end{equation}
where the effective gravitational mass $M_G(r)$ is defined by
\begin{equation}
M_G(r)=\frac{1}{2}re^{\frac{\mu-\nu}{2}\nu'}.
\end{equation}
Substituting Eq. (46) into Eq. (45), we obtain the simple expression
\begin{equation}
-\frac{\nu'}{2}(\rho +p_{r})-\frac{dp_{r}}{dr}+\frac{2}{r}(p_{t}-p_{r})=0.
\end{equation}

Summarising,  this expression is arranged  in terms of gravitational mass and hence
 gives the equilibrium condition for the compact star, involving the gravitational,
hydrostatic and anisotropic forces for stellar objects.  This equation  can be expressed in the more compact form as
\begin{equation}
F_{g}+F_{h}+F_{a}=0,
\end{equation}
where $F_g =-\frac{\nu'}{2}(\rho+p_r)$, $F_h =-\frac{dp_r}{dr}$ and $F_a=\frac{2}{r}(p_t-p_r)$
represents the gravitational, hydrostatic and anisotropic forces, respectively.
In order to illustrate this qualitatively, we use graphical representation for both cases
when f(R) = R and f(R) = aR+b, which are shown in Figs. 3, 5 and 7 (right panel)
by assigning the value of $\omega$ = 0.58. In spite of this model we see that
in every cases $F_h$ and $F_a$  takes the negative value while $F_g$ is positive.
As a result, it is clear that gravitational force is counterbalanced by the combined effect of hydrostatic
and anisotropic forces to hold the system in static equilibrium. Models to explain
the static equilibrium in-depth have been extensively studied by
Rahaman et al. \cite{rah16} and  Rani \& Jawad \cite{jawad}.

\subsection{Stability Analysis}
We are interested in checking the  sound speed, using the concept of Herrera's
cracking (or overtuning) \cite{Herrera}, which lies within the range $0 < v_{si}^{2}\leq 1$
i.e., according to this procedure the radial and transverse velocity of sound lies
within the proposed range. We have shown graphically
the radial and transverse velocity of sound for both cases when f(R) = R and f(R) = aR + b,
which are shown in Figs. 3, 5 and 7 (left panel) and observe that these parameters satisfy the inequalities
$0 < v_{sr}^{2}\leq 1$ and $0 < v_{st}^{2}\leq 1$ everywhere within the stellar object
except the isotropic case when f(R) = R.

\section{Conclusion}

In this work, we have studied analytical solutions for compact
stellar objects with a general static interior source in the framework of f(R) gravity
satisfying the conformal Killing vectors equations.
The investigation has been performed for two specific cases when f(R) = R and
f(R) = aR + b; further the stars are assumed to be anisotropic in their internal structure.
In order to get exact solutions we have considered a systematic approach by assuming the
spherical symmetry and interior of the dense star admitting non-static conformal symmetry.

For a simplest form of the fluid sphere, we have studied the two
specific arguments for both isotropic $(p_{r} = p_{t})$ as well as non isotropic $(p_{r}\neq p_{t})$ cases.
Interestingly, we have identified that among the two cases ( case 4.1 and 4.2 ), only
the solution for isotropic condition $p_{r} = p_{t} = p$ when f(R) = R is not physically valid
because the behavior of different energy conditions does not attribute the regularity conditions
at the interior of the star. Furthermore, it has been found that our solution satisfies
all the energy conditions and that  pressures are positive and finite throughout interior of the stars
which are needed for physically possible configurations have been discussed in detail.
For a stable configuration, we have shown that the generalized TOV equation which
describes the equilibrium condition for an anisotropic fluid subject to by the different
forces, viz. gravitational force ($F_g$), hydrostatic force ($F_h$) and anisotropic force
($F_a$) in \textbf{ Figs. 3, 5 and 7} (left panel). Another interesting result
of this paper is related with checking the stability of our model by adapting
Herrera's cracking concept \cite{Herrera} and it has been found that the radial
and transverse speeds of sound lies within the limit of $(0, 1]$ which are shown
in \textbf{ Figs. 3, 5 and 7} (reght panel). Hence we concluded that our solution
might have astrophysical relevance by this theory through fine tuning.
Therefore, it would be an interesting task to verify our solution with sample
data for more satisfactory features in the realm of physical reality which will
be our next venture in this line of study. Moreover, we propose to consider  more general functional forms of $f$ to go beyond the standard model of general relativity as considered in this work.

\subsection*{Acknowledgments}
 AB, SB are thankful to the authorities of Inter University Centre
for Astronomy and Astrophysics, Pune for giving them an opportunity to visit IUCAA
 where a part of this work was carried out.

\end{document}